\documentclass[aps,prc,showpacs,twocolumn,superscriptaddress,10pt]{revtex4-1}
\usepackage{amsmath}
\usepackage{amssymb}
\usepackage{graphicx}
\begin{document}
\title{Analysis of the total kinetic energy of fission fragments with the Langevin equation}
\author{M.D. Usang}
\email{usang.m.aa@m.titech.ac.jp}
\email{mark\_dennis@nm.gov.my}
\affiliation{Laboratory for Advanced Nuclear Energy, Institute of Innovative Research, Tokyo Institute of Technology, 2 chome-12-1 Ookayama, Meguro, Tokyo 152-8550, Japan.}
\affiliation{Reactor Technology Center, Technical Support Division, Malaysia Nuclear Agency, Bangi, 43000 Kajang, Selangor Darul Ehsan, Malaysia.}
\author{F.A. Ivanyuk}
\affiliation{Laboratory for Advanced Nuclear Energy, Institute of Innovative Research, Tokyo Institute of Technology, 2 chome-12-1 Ookayama, Meguro, Tokyo 152-8550, Japan.}
\affiliation{Nuclear Theory Department, Institute for Nuclear Research, Prospect Nauki 47, 03028 Kiev, Ukraine.}
\author{C. Ishizuka}
\affiliation{Laboratory for Advanced Nuclear Energy, Institute of Innovative Research, Tokyo Institute of Technology, 2 chome-12-1 Ookayama, Meguro, Tokyo 152-8550, Japan.}
\author{S. Chiba}
\affiliation{Laboratory for Advanced Nuclear Energy, Institute of Innovative Research, Tokyo Institute of Technology, 2 chome-12-1 Ookayama, Meguro, Tokyo 152-8550, Japan.}
\affiliation{Theoretical Division, National Astronomical Observatory of Japan, 2 chome-21-1 Osawa, Mitaka, Tokyo 181-0015, Japan.}

\date{\today}

\begin{abstract}
We analyzed the total kinetic energy (TKE) of fission fragments with three-dimensional Langevin calculations for a series of actinides and Fm isotopes at various excitation energies. This allowed us to establish systematic trends of TKE with $Z^2/A^{1/3}$ of the fissioning system and as a function of excitation energy.  In the mass-energy distributions of fission fragments we see the contributions from the standard, super-long and super-short (in the case of $^{258}$Fm) fission modes.
For the fission fragments mass distribution of $^{258}$Fm we obtained a single peak mass distribution.
The decomposition of TKE into the prescission kinetic energy and Coulomb repulsion showed that decrease of TKE with growing  excitation energy is accompanied by a decrease of prescission kinetic energy.  It was also found that transport coefficients (friction and inertia tensors) calculated by a microscopic model and by macroscopic models give drastically different behavior of TKE as a function of excitation energy. The results obtained with microscopic transport coefficients are much closer to experimental data  than those calculated with macroscopic ones.

\end{abstract}

\maketitle

\section{Introduction}
The nuclear fission phenomena is very fascinating because it involves a large-scale restructuring of nucleon arrangements. The motion of each individual nucleon can be taken into account in approaches such as time-dependent Hartree-Fock theory or molecular dynamics that consider the degrees of  freedom of all nucleons in the system quantum-mechanically. However, it is not possible yet to treat the nuclear fission starting from the compound nuclei all the way to scission by these microscopic theories.  On the other hand, in the Langevin description of fission, 
we keep only a small number of collective coordinates which are convenient to describe nuclear fission assuming that the time-evolution of the collective shape of the nucleon distribution can be described by the classical treatment.
The Langevin approach extends the classical Newtonian equation by adding a random force. In nuclear fission, the random force is due to the sum of fluctuations resulting from the complex changes of each individual nucleons movements acting on the collective coordinates.

We can describe all the possible evolution of the nuclear shape by these collective coordinates. Given a particular set of initial collective coordinates $q_{\mu}$ for $\mu=\{1,..,D$\} where $D$ denotes the number of collective coordinates, we then allow the shape of the nuclei to evolve on the potential energy surface under the influence of friction and the random force with trajectories determined by the Langevin equation. We let the shape evolve all the way to scission configurations if features necessary for fission are present on the potential energy surface.

At present there are several groups using the Langevin approach for the description of fission processes \cite{ mazurek2017, eslamizadeh2017, usang2016, pahlavani2015, mazurek2015, asano2004,Sierk2017}.
In all these works only potential energy is calculated accurately enough mainly within the macroscopic-microscopic method which combines liquid-drop properties of fissioning nuclei with quantum shell and pairing effects. The tensors of friction and inertia are calculated within macroscopic models: the Werner-Wheeler method \cite{massliquid} for the inertia and the wall-and-window formula \cite{wallfriction,adeev,krapom,swiat1984,nixsierk} for friction. These quantities do not contain any quantum effects. Meanwhile, it was shown \cite{ivhopaya,ivahof} that the mass and friction coefficients derived within a microscopic approach at low excitation energies differ drastically from their macroscopic counterparts in dependence on both the deformation and excitation energy (temperature). Thus, in nuclear fission at low excitations the application of macroscopic transport coefficients is not well justified.

\begin{figure}[b]
\centering
\includegraphics[width=0.3\textwidth]{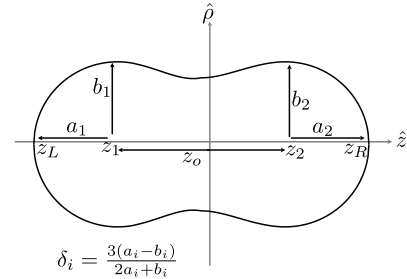}
\caption{\label{profile} The two center shell model shape profile.}
\end{figure}

In the present work, we  describe the mass-energy distributions of actinide nuclei and $^{258}$Fm within  the three-dimensional (3D) Langevin approach with transport coefficients derived within the microscopic linear response theory \cite{hofmann1997q,ivanyuk2000}.  The shape of the nuclear surface and the mean-field Hamiltonian are defined within the two center shell model \cite{MaruhnGreiner1972} with three deformation parameters,  
namely, the elongation $z_0$, the deformation of fragments $\delta$, and the mass asymmetry $\alpha$ that specify the nuclear shape as depicted in Fig. \ref{profile} during the fission process.


The total kinetic energy (TKE) of fission fragments is an important fission observables as it reveals further information on excitation energy ($E_x$) distribution of fission fragments for the calculation of prompt fission neutron multiplicities \cite{capote2016pfns}.
Presently, TKE at various $E_x$ are predicted mainly from available data \cite{madland2006,king2017}, but our approach opens possibilities of making reasonable TKE estimate for gaps in available experimental data.
Calculations using macroscopic transport coefficients are also carried out for comparison in some cases.
\section{The two-center shell model}
\label{TCSM}
In the present  work we use the two-center shell model (TCSM) suggested by Maruhn and Greiner \cite{MaruhnGreiner1972} and the code developed by Suekane, Iwamoto, Yamaji and Harada \cite{suek74,iwam76,sato79} and extended by one of the authors (Ivanyuk).

The mean-field Hamiltonian $H_{mf}$ in TCSM includes the kinetic energy part, the mean-field potential $V(\rho, z)$, and the angular momentum dependent part.
In cylindrical coordinates $\{\rho, z\}$ it is written as
\begin{equation}\label{hamil}
H_{mf}=\frac{\vec p~^2}{2m} + V(\rho, z)-\kappa_i[2(\vec l_i \vec s)+\mu_i (\vec l_i^2-\langle\vec l_i~^2\rangle)]\hbar\omega_0;
\end{equation}
see \cite{suek74,iwam76,sato79}. Here $i=1$ for $z\leq 0$ and $i=2$ for $z\geq 0$; $\kappa_i$ and $\mu_i$ are the usual parameters of Nilsson model.

\begin{figure}[ht]
\centering
\includegraphics[width=0.4\textwidth]{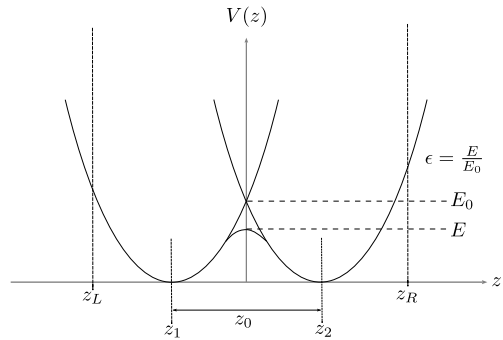}
\caption{\label{potential} The nuclear potential \protect\eqref{v_tcsm} of the two-center shell model at $\rho=0$.}
\label{fig1}
\end{figure}

The potential $V(\rho, z)$ in TCSM consists of two oscillator potentials  smoothly joined together by a fourth-order polynomial in $z$; see Eq. \eqref{v_tcsm} and Fig. \ref{fig1}.  It is defined as,
\begin{equation}\label{v_tcsm}
  V(\rho,z) = \left\{
 \begin{array}{lr}
 \frac{1}{2}m \omega_{z_1}^2 (z-z_1)^2 + \frac{1}{2}m \omega_{\rho_1}^2 \rho^2,& z \leq z_1 \\
 \frac{1}{2}m \omega_{z_1}^2 (z-z_1)^2 f_1(z,z_1) + \\
 \frac{1}{2}m \omega_{\rho_1}^2 \rho^2 f_2(z,z_1),&  z_1 \leq z \leq 0 \\
 \frac{1}{2}m \omega_{z_2}^2 (z-z_2)^2 f_1(z,z_2) + \\
 \frac{1}{2}m \omega_{\rho_2}^2 \rho^2 f_2(z,z_2),&  0 \leq z \leq z_2 \\
 \frac{1}{2}m \omega_{z_2}^2 (z-z_2)^2 + \frac{1}{2}m \omega_{\rho_2}^2 \rho^2,& z_2 \leq z
 \end{array}\right.
\end{equation}
with the quadratic-in-$z$ functions $f_1$ and $f_2$,
\begin{align}\label{f1f2}
f_1(z,z_i) =& 1 + c_i (z-z_i) + d_i (z-z_i)^2 ;\nonumber \\
f_2(z,z_i) =& 1 + g_i (z-z_i)^2,~~(i=1,2) .
\end{align}

The shape of nuclear surface in TCSM is fixed by the requirement that at the surface $\rho=\rho(z)$ the potential  $V(\rho(z),z)$ is constant, which leads to
\begin{equation}\label{rho_tcsm}
  \rho^2(z) = \left\{
 \begin{array}{lr}
 \frac{\omega_{z_1}^2}{\omega_{\rho_1}^2}  [(z_L-z_1)^2 - (z-z_1)^2];& z_L\leq z \leq z_1 \\
 \frac{\omega_{z_1}^2}{\omega_{\rho_1}^2}\frac{(z_L-z_1)^2 - (z-z_1)^2f_1(z,z_1)}{f_2(z,z_1)};&  z_1 \leq z \leq 0 \\
 \frac{\omega_{z_2}^2}{\omega_{\rho_2}^2}\frac{(z_R-z_2)^2 - (z-z_2)^2f_1(z,z_2)}{f_2(z,z_2)};&  0 \leq z \leq z_2 \\
 \frac{\omega_{z_2}^2}{\omega_{\rho_2}^2}  [(z_R-z_2)^2 - (z-z_2)^2];& z_2 \leq z\leq z_R
 \end{array}\right.
\end{equation}
The potential \eqref{v_tcsm} contains 12 parameters. By imposing conditions that
$V(\rho, z)$ and its $z$ derivative are continuous at $z=\{z_1, 0, z_2\}$, the number of parameters is reduced to 5 (one also has to require that the left and right oscillator potentials coincide at $z=0$). These are the elongation parameter $z_0\equiv z_2-z_1$, the mass asymmetry $\alpha={(V_1 - V_2)}/{(V_1 + V_2)}$ ($V_1$ and $V_2$ are the volume to the left and right from $z=0$), the deformations $\delta_i$ of the left and right oscillator potentials, and the neck parameter $\epsilon$.

The neck parameter $\epsilon$ is given by the ratio of the
potential height $E$ at $z=0$ to the value $E_{0}$ of left and right harmonic oscillator potentials at $z=0$, see Fig~\ref{v_tcsm}.
In our calculation, we fix $\epsilon=0.35$ as was recommended by \cite{yama88}. This value leads to shapes that are very close to the optimal shapes \cite{ivchar}. Please, note that the fixed $\epsilon$ does not mean fixed neck radius. The neck radius depends on all five deformation parameters. For fixed $\epsilon$, we can already see that the variation of elongation parameter $z_0$ alone leads to a very reasonable sequence of shapes of the fissioning nucleus, see \citet{aritomo2014}.

The two additional parameters $z_L$ and $z_R$ that appear in \eqref{rho_tcsm} are the left and right tips of the nucleus, $\rho(z_L)=\rho(z_R)=0$. They can be found from the conditions,  $\pi\int^{z_L}_0 \rho^2(z) dz=V_1\,$ and $\, \pi \int_0^{z_R} \rho^2(z) dz=V_2$.

All parameters that appear in Eq. \eqref{v_tcsm}--\eqref{rho_tcsm} can be expressed in terms of above five deformation parameters, for example,
\begin{eqnarray}\label{tcsm_params}
{\omega_{z_i}}&/&{\omega_{\rho_i}} = {(3-2 \delta_i)}/{(3+\delta_i)}\equiv\beta_i, \nonumber\\
g_1&=&\frac{1-Q^2}{Q^2(1+Q\,\beta_1/\beta_2)},\,\,
g_2=\frac{Q(Q^2-1)\,\beta_1/\beta_2}{(1+Q\,\beta_1/\beta_2)}\nonumber\\
c_1&=&c_2=2-4\epsilon,\quad d_1=d_2=1-3\epsilon.
\end{eqnarray}
The quantity $Q\equiv \omega_{\rho_1}/\omega_{\rho_2}$ should be found numerically from the volume conservation condition $\int_{z_L}^{z_R}\rho^2(z)\,dz=(4/3)R_0^3$, where $R_0$ is the radius of the spherical compound nucleus, $R_0=r_0\,A^{1/3}$. For $r_0$ we use the value $r_0=1.2$ fm.

As one can see from Eq. \eqref{rho_tcsm}, the ratio $\omega_{z_i}/\omega_{\rho_i}$ and, thus, the deformation parameters $\delta_i$ (or $\beta_i$) represent only the ratio  of semi-axes of the outer ($z_L\leq z\leq z_1$ or $z_2\leq z\leq z_R$) parts of the shape of nucleus.
Therefore, in general, it does not mean that the fragments are prolate if $\delta_i$ is positive, or oblate when $\delta_i$ is negative.

In the present work we perform the Langevin calculations in the three-dimensional space of deformation parameters $\{q\}=\{z_0/R_0, \; \delta, \; \alpha\}$. That is, we impose a constraint such that $\delta_1=\delta_2\equiv \delta$.
In this case  $\beta_1=\beta_2\equiv \beta_0$.

The known deficiency in the shape parametrization of the two-center shell model at small $z_0$ and finite $\alpha$ is discussed and resolved in \cite{usang2016} by imposing the condition $\rho(z=z_1)=\rho(z=z_2)$ and introducing $\alpha$-dependent deformation parameters,
\begin{subequations}
\label{smooth}
\begin{align}
\beta_1(\alpha)=&f(z_0)\beta_0[1+\alpha]+[1-f(z_0)]\beta_0, \label{smooth1} \\
\beta_2(\alpha)=&f(z_0)\beta_0[1-\alpha]+[1-f(z_0)]\beta_0, \label{smooth2}
\end{align}
\end{subequations}
with $f(z_0)=[1+\exp((z_0-R_0)/(0.2R_0))]^{-1}$. The factor $f(z_0)$ ensures that the $\alpha$-dependence of $\beta_i$ is effective only for small values of $z_0$. For large values of $z_0$, $\beta_i(\alpha)$ turn into $\beta_0$. In present work the scission dynamics is governed by a part of the potential energy
surface where $\delta_1 = \delta_2$.  Removal of this constraint is ongoing in our group as will be described later.
\section{The Langevin approach}
\label{lange}
The Langevin equations form a system of first-order differential equations for the collective coordinates $\{q_\mu\}$ and their conjugate momenta $\{p_\mu\}$ \cite{abe} describing the time evolution of the collective coordinates.
It is succinctly written as,
\begin{eqnarray}\label{langevin}
\frac{dq_\mu}{dt}&=&\left(m^{-1} \right)_{\mu \nu} p_\nu  ,\\
\frac{dp_\mu}{dt}&=&-\frac{\partial U(q)}{\partial q_\mu} - \frac{1}{2}\frac{\partial m^{-1} _{\nu \sigma} }{\partial q_\mu} p_\nu p_\sigma
 -\gamma_{\mu \nu} m^{-1}_{\nu \sigma} p_\sigma +g_{\mu \nu} R_\nu (t), \nonumber
\end{eqnarray}
where summation over repeated indices is assumed.
The potential energy surface $U(q)$ along which the shape evolves according to Eqs. \eqref{langevin} is calculated by the macroscopic-microscopic approach \cite{strut67,strut68,brdapa},
\begin{equation}
\label{vdef}
U(q)=E^{LD}_{def}(q)+\delta E(q).
\end{equation}
The potential energy surface $U(q)$ contains the  contributions from the liquid drop deformation energy, $E^{LD}_{def}(q)=E_{LD}(q) - E^{sph}_{LD}$ and from the shell and pairing corrections,
\begin{equation}
\label{shell-pair-corr}
\delta E(q) =\sum_{n,p} \left( \delta E_{shell}^{(n,p)}(q) + \delta E_{pair}^{(n,p)} (q) \right) .
\end{equation}
The summation in \eqref{vdef} is carried out over the neutrons (n) and protons (p).
The correction, $\delta E^{(n,p)}_{shell}(q)$ is calculated as the difference between the sum of single particle energies of the occupied states and its averaged value is defined by Strutinsky smoothing.  The single-particle energies are calculated with the two-center shell model \cite{MaruhnGreiner1972,suek74,iwam76,sato79}. The pairing interaction was taken into account by  BCS approximation and the shell correction to the pairing correlation energy, $\delta E^{(n,p)}_{pair}(q)$, was evaluated by the method suggested in \cite{brdapa}.

We consider in present work the fission process at low excitation energies. The corresponding temperatures  do not exceed $1$ MeV. For such temperatures the damping of shell effects is not so large and it was neglected. So, the calculations were done with full shell effects taken into account.

The liquid drop energy is obtained from the finite-range liquid drop model \cite{krap79} as the sum of surface energy $E_S$ and Coulomb energy $E_C$. In the code \cite{suek74,iwam76,sato79} the
following parameters of the finite-range liquid drop model \cite{krap79}
were used:  $r_{0}=1.20$ fm, $a=0.65$ fm, $a_{s}=21.836$ MeV, and
$\kappa_{s}=3.48$, where $r_{0}$ and $a$ are the nuclear-radius
constant and the range of the Yukawa folding function, and $a_{s}$
and $\kappa_{s}$ are the surface energy constant and the
surface asymmetry constant, respectively.
The liquid drop energy for the spherical shape, $E^{sph}_{LD}$, is obtained in the same manner but for a spherical nuclei.

The main results in this work were obtained with the so-called microscopic transport coefficients \cite{usang2016}: the mass and friction tensors were calculated using linear response theory and a locally harmonic approximation \cite{hofmann1997q}. 
The linear response function are further explained in Sec. IV.
The exact expressions for the mass and friction tensor at finite temperatures for a system with  pairing  can be found in \cite{ivanyuk2000}. The mass $m_{\mu \nu}$ and friction $\gamma_{\mu \nu}$ tensors were calculated at a fixed grid points in temperature. The values for the actual (local) temperature were found by the interpolation between grid points.

In some cases, for comparison, we carried out also the Langevin calculation with macroscopic transport coefficients.  The macroscopic transport coefficients were calculated in a standard way. We used the Werner-Wheeler method \cite{massliquid} for the inertia tensor and the wall-and-window formula \cite{wallfriction,adeev,krapom,swiat1984,nixsierk} for the friction tensor.

The random force in Eq. \eqref{langevin} is written as the product of white noise $R_\nu$ and the strength factors $g_{\mu \nu}$.
The strength factors $g_{\mu \nu}$ are
expressed in terms of the diffusion tensor $D_{\mu \nu}$,
$D_{\mu \nu}=g_{\mu \sigma}g_{\nu \sigma}$, which is
related to the friction tensor $\gamma_{\mu \nu}$ via the
modified Einstein relation,
\begin{equation}
\label{Einstein relation}
D_{\mu \nu}=T^* \gamma_{\mu \nu}=g_{\mu \sigma} g_{\nu \sigma}.
\end{equation}

Here $T^*$ is the effective temperature introduced by Hofmann \cite{hofngo,hofmann-low-temp},
\begin{equation}\label{tstar}
T^* = \frac{\hbar\varpi}{2} \coth \frac{\hbar\varpi}{2T}\,.
\end{equation}
The point is that the classical Einstein relation $D=T \gamma$ is valid
at relatively high temperatures.  At low temperatures the quantal aspect of the
fluctuation-dissipation theorem becomes important and the magnitude of the diffusion
coefficient becomes larger than its classical value.

This property is guaranteed by the form (\ref{tstar}). Here parameter $\varpi$ is
the local frequency of collective motion. In principle, it should be calculated at each deformation point.
Unfortunately, this would be too time consuming.

 The minimum of $\hbar\varpi$
is given by the zero-point energy. Based on the pioneer works \cite{Bohr+1939,Hill+1952}, we estimated the zero-point energy in the 3D case to be $\hbar\varpi\approx 0.4\, \rm{MeV} \times 3= 1.2 \, \rm{MeV}$,
because in the 1D case the zero-point energy was considered to be equal to 0.4 MeV. Since the zero-point energy represents only the minimal value of $\hbar\varpi$, we used for $\hbar\varpi$ the somewhat larger value $\hbar\varpi$=2 MeV independently of deformation. In this case $T^*=$1~MeV at $T=0$. In the high-temperature limit $T^*$ coincides with $T$.
The variation of $\hbar\varpi$ within reasonable limits does not change much the calculated distributions of fission fragments.

The temperature $T$ was related to the thermal intrinsic energy $E_{int}$ by the Fermi  gas relation,
\begin{equation}
\label{Fermi-gas}
E_{int}=aT^2 ,
\end{equation}
where $a$ denotes the level density parameter \cite{level-density},
\begin{equation}
a=A\,[1.0+3.114\,A^{-1/3}+5.626\,A^{-2/3}]\,/\,14.61\,.
\end{equation}
The $E_{int}$ in \eqref{Fermi-gas} is the intrinsic excitation energy, calculated at each step of the integration of equations of motion from the energy balance,
\begin{equation}
\label{eint}
E_{int}=E_x -\frac{1}{2}\left [m^{-1}(q) \right ]_{\mu\nu}p_{\mu}p_{\nu}-U(q),
\end{equation}
where $E_x$ is the initial excitation energy of the system. For neutron induced fission $E_x=S_n + E_n$, where $S_n$ and $E_n$ are the neutron separation energy and incident neutron kinetic energy, respectively.

\section{Transport Coefficients for Dynamical Calculations}
\label{transport}
\subsection{Macroscopic transport coefficients}
In the present work we also show the so called {\it macroscopic transport coefficient} which are often used to solve the Langevin equation. The macroscopic transport coefficients depend only on the shape (deformation) of the system. They do not contain any quantum effects, nor the dependence on the excitation energy (temperature of the system).
The macroscopic mass tensor $M_{\mu\nu}^{WW}$ is usually defined in the Werner-Wheeler approximation \cite{massliquid},
\begin{equation}
M_{\mu\nu}^{WW} =
 \pi \rho_{_0} \int_{z_L}^{z_R} \rho^2 \left[ A_{\mu} A_{\nu}  
+ \frac{\rho^2}{8} A_{\mu}' A_{\nu}' \right] dz,
\end{equation}
where $\rho$ is a function with respect to $z$ and 
\begin{equation}
A_{\mu}(z;q)=\frac{1}{\rho^2(z,q)}\frac{\partial}{\partial q_{\mu}} \int_{z}^{z_{_R}} \rho^2 (z',q) dz'.
\end{equation}
The corresponding macroscopic friction tensor is the so-called wall-and-window formula \cite{wallfriction,swiat1984,nixsierk}. 
The loss of collective energy is given by $\dot{E}=(3/4)\rho_{_0} v_F \oint v_n^2 (s) ds$, where $v_F$ and $v_n$ are the the Fermi velocity and normal velocity of the nuclear surface respectively. 
The definition of nuclear density is $\rho_{_0}=A/(4\pi R_0^3/3)$. 
The Fermi velocity was estimated from the relationship, $\hbar k_F=m v_F$ and the Fermi momentum was estimated from the Fermi gas, $k_F R_{_0}=\sqrt[3]{9\pi A/4}$.
According to \citet{wallfriction}, the wall friction coefficient, $\gamma^{\mathrm{wall}}$ is proportional to $v_n^2(s)$, and it was suggested that $\dot{E}=\sum_{\mu\nu} \gamma_{\mu\nu}^{\mathrm{wall}} \dot{q}_{\mu} \dot{q}_{\nu}$. 
In the case of axial symmetric shapes, $v_n(s)$ can be expressed in terms of the profile function $\rho(z;q)$, and the wall friction can be written as
\begin{equation}\label{frwall}
\gamma_{\mu\nu}^{\mathrm{wall}}=\pi\rho_{_0} v_F\int_{z_{_L}}^{z_{_R}}dz \,\frac{\partial \rho^2}{\partial q_{\mu}}\frac{\partial \rho^2}{\partial q_{\nu}}
\left[4\rho^2+\left(\frac{\partial\rho^2}{\partial z} \right)^2 \right]^{-1/2}_.
\end{equation}

As the nucleus begins to separate, it is necessary to describe the wall friction as the sum of both the left and right fragments' friction with respect to the center-of-mass velocities of the fragments \cite{adeev,krapom},
\begin{equation}\label{frwall_2}
\gamma_{\mu\nu}^{\mathrm{wall2}}=\frac{\pi\rho_{_0}\bar v}{2}\left(\int_{z_{L}}^{0}I_L(z)\,dz+\int_{0}^{z_{_R}}I_R(z)dz \right)_,
\end{equation}
with
\begin{align}\label{ILR}
&I_{L,R}(z)=
\left(\frac{\partial \rho^2}{\partial q_{\mu}}+\frac{\partial \rho^2}{\partial z}\frac{\partial z_{cm}(L,R)}{\partial q_{\mu}}\right) \times \nonumber \\
&\left(\frac{\partial \rho^2}{\partial q_{\nu}}+\frac{\partial \rho^2}{\partial z}\frac{\partial z_{cm}(L,R)}{\partial q_{\nu}}\right)
\left[4\rho^2+\left(\frac{\partial\rho^2}{\partial z} \right)^2 \right]^{-1/2}_.
\end{align}

\citet{swiat1984} suggests that, as the shape comes closer to a scission configuration, a correction called the \textit{window} terms could be introduced, where nucleons from the left fragment can traverse through the window into right fragment and vice versa, as well collision between the nucleons that tries to traverse this window.
Given that the volume and the center-of-mass distance between the two fragments, $R_{12}$, are moving with respect to time, the \textit{window} friction term can be written as
\begin{equation}
\gamma_{\mu\nu}^{\mathrm{window}}=\frac{\rho_{_0}\bar{v}}{2} \left[ \Delta \sigma   
\frac{\partial R_{12}}{\partial q_\mu} \frac{\partial R_{12}}{\partial q_\nu} + \frac{32}{9\Delta \sigma} \frac{\partial V_L}{\partial q_{\mu}}  \frac{\partial V_L}{\partial q_{\nu}}\right]_,
\end{equation}
giving us the \textit{wall-window} friction tensor,
\begin{equation}
\gamma_{\mu\nu}^{w+w}=\gamma_{\mu\nu}^{\mathrm{wall2}} +\gamma_{\mu\nu}^{\mathrm{window}}.
\end{equation}
The transition between the regime governed by $\gamma^{\mathrm{wall}}$ and $\gamma^{w+w}$ should be smooth, leading to the phenomenological ansatz proposed by \citet{nixsierk},
\begin{equation}
\gamma_{\mu\nu}^{\mathrm{total}}=\sin^2(\pi \alpha/2)\gamma_{\mu\nu}^{\mathrm{wall}} + \cos^2(\pi \alpha /2) \gamma_{\mu\nu}^{\mathrm{w+w}},
\end{equation}
determined using the condition $\alpha=(r_{\mathrm{neck}}/R_{\mathrm{min}})^2$. $R_{\mathrm{min}}$ is the minimal semi-axis of two outer ellipsoids in a three-quadratic-surfaces shape parametrization and $r_{\mathrm{neck}}$ is the radius of the neck.
Usually, this friction is too large, so we multiply this friction with the shape independent reduction factor, $k_s=0.27$.

In most cases, these macroscopic transport coefficients are able to reproduce a decent approximation of the actual friction experienced during fission.
The drawback of macroscopic transport coefficients is their independence with regards to temperature, and this drawback can be observed in the results that we will give later.

\subsection{Microscopic transport coefficients}
In the linear response approach to the nuclear collective motion \cite{hofmann1997q} the nuclear many-body Hamiltonian $\hat H$ is represented as the sum of the deformed (time dependent) mean field Hamiltonian $\hat H_{mf}(q(t))$ and the residual interaction $\hat V^{(2)}_{res}$  which is assumed to be deformation independent, $\hat H=\hat H_{mf}+\hat V^{(2)}_{res}$. In this case the derivative of Hamiltonian $\hat H$ with respect to deformation $q$
\begin{equation}\label{F}
\hat{F} ( q) \equiv\frac{\partial \hat H_{mf} (q)}{\partial q}\,,
\end{equation}
is a one-body operator. The total energy of the system is given by the mean value of $\hat H$, $E_{tot}=\langle \hat H \rangle_t$. By the index $t$ we have indicated that the average $<\,\,\,>_t$ has to be  calculated
with a time-dependent, nonequilibrium density operator. The equation of motion for $q(t)$ can be constructed from the energy conservation condition. Differentiating $E_{tot}$ with respect to $t$, one gets from Ehrenfest's theorem
\begin{eqnarray}\label{Ehrenfest}
\frac{dE_{tot}}{dt}&=&\frac{d\langle \hat H \rangle_t}{dt} =\frac{d\langle \hat H_{mf} (q) \rangle_t}{dt} =\frac{}{}\langle \hat F(q) \rangle_t \dot q = 0\,,\quad\nonumber\\
&& \Rightarrow \langle \hat F(q) \rangle_t=0.
\end{eqnarray}
What is left is to express $\langle \hat F(q) \rangle_t$ as functional  of $q(t)$. For this purpose let us expand $\hat F(q)$ around some point $q_0$,
\begin{eqnarray}\label{expand}
\hat F(q)&=&\hat F(q_0)+(q-q_0)\left\langle {d \hat F}/{dq_0} \right\rangle_{q_0} \nonumber\\
&=&\hat F(q_0)+(q-q_0)\left\langle {d^2 \hat H}/{d q_0^2} \right\rangle_{q_0}\,,
\end{eqnarray}
and estimate $\langle \hat F(q_0) \rangle_t$ by perturbation theory up to the linear order in $q(t)-q_0$:
\begin{equation}\label{favr}
\Delta\langle \hat F \rangle_t\equiv\langle \hat F(q_0) \rangle_t-\langle \hat F(q_0) \rangle_{q_0}=-\int
\widetilde\chi(t-s) (q(s) - q_0)ds 
\end{equation}
The index $q_0$ in Eqs.(\ref{expand}) and (\ref{favr}) means that the average $<\,\,\,>_{q_0}$ has to be
calculated with the quasi-static density matrix at $q=q_0$.
The quantity $\widetilde\chi$ in (\ref{favr}) is the causal response function \cite{hofmann1997q}
\begin{equation}\label{chit}
\widetilde\chi(t-s)=\Theta(t-s)\frac{i}{\hbar}\langle [\hat F^I(t),\hat F^I(s)] \rangle_{q_0}.
\end{equation}
Here $\Theta(t-s)$ is  the step function, $\Theta(t-s)$ equals one if $t\ge s$ and zero elsewhere.
The time dependence of $\hat F^I(t)$ (in interaction representation) is defined by $\hat H_{mf}(q_0)$, i.e., by the properties of the system at $q_0$,
\begin{equation}\label{fint}
\hat F^I(t)=e^{-\frac{i}{\hbar}\hat H_{mf}(q_0)t}\hat F(q_0)e^{\frac{i}{\hbar}\hat H_{mf}(q_0)t}.
\end{equation}
In the representation of eigenfunctions of $\hat H_{mf}(q_0)$, $\widetilde\chi(t)$ takes the form
\begin{equation}\label{chitkj}
\widetilde\chi(t)=\sum_{jk} ( n_k - n_j)F_{jk} F_{kj} e^{\frac{i}{\hbar}(\varepsilon_k - \varepsilon_j  + i\epsilon)t},
\end{equation}
where subscripts $k$ and $j$ are indices of single-particle states, $\epsilon_k$ is the single particle energy, and $n_k=1/[1+\exp((\epsilon_k-\lambda)/T)]$.  The symbol $F_{jk}$ denotes a matrix element of $\hat{F}$ between states $j$ and $k$.
The infinitely small term $i\epsilon$ in the exponent in (\ref{chitkj}) appears due to the assumption of infinitely slow switching on of the perturbation.
Inserting (\ref{favr}) and (\ref{chit}) into (\ref{Ehrenfest}), one comes to the equation
\begin{equation}\label{eom1}
-\int_{-\infty}^{\infty}\widetilde\chi(t-s) \Delta q(s)ds+\Delta q(t)\,\left<{d^2 \hat{H}}/{dq^2} \right>_{q_0}=0,
\end{equation}
with $\Delta q(s)\equiv q(s)-q_m$.
Here $q_m$ is the point where $\langle \hat F(q) \rangle_{q}$ turns into zero. It is defined by the equation
\begin{equation}\label{qm}
\langle \hat F(q_0) \rangle_{q_0}+(q_m-q_0)\langle{\partial^2 \hat H}/{\partial q_0^2}\rangle_{q_0}=0.
\end{equation}
Taking the Fourier transform of (\ref{eom1}), one gets the secular equation
\begin{equation}\label{secular}
\chi  \left( \omega \right) +  k^{-1}= 0 \,,
\end{equation}
with the coupling constant $k$ given by
\begin{equation}\label{couple}
-k^{-1} = \left<{d^2 \hat{H}}/{dq^2} \right>_{q_0} \ .
\end{equation}
The Fourier transform of $\widetilde\chi(t)$ is
\begin{equation}\label{chiom}
\chi(\omega)=\int_{-\infty}^{\infty}\widetilde\chi(t)e^{i\omega t}\,dt
= \sum_{jk} \frac{(n_k - n_j)F_{jk} F_{kj}}{\hbar \omega - \left(\varepsilon_k - \varepsilon_j \right) + i\epsilon}  \ .
\end{equation}
From the definition (\ref{couple}) of the coupling constant it follows that it can be expressed as the sum of the   static response $\chi(\omega=0)$ and the stiffness $C(0)$ of the potential energy
\begin{equation}\label{couple2}
-k^{-1}=\chi(\omega=0)+C(0),\,\text{with}\, C(0)=\frac{d^2\langle \hat H_{mf}\rangle}{d q_0^2} .
\end{equation}

In the case of slow collective motion one can solve the secular equation (\ref{secular}) by expanding $\chi(\omega)$ up to the second order in powers of $\omega$:
\begin{equation}\label{seqular2}
\chi(\omega=0) + \omega \left.\frac{d\chi}{d\omega}\right|_{\omega=0}+ \frac{1}{2}\omega^2 \left.\frac{d^2\chi}{d\omega^2}\right|_{\omega=0} +  k^{-1}= 0 \ .
\end{equation}
By  multiplying (\ref{seqular2}) with $\Delta q(\omega)$ and performing the inverse Fourier transform, one gets the equation of motion for an oscillator:
\begin{equation}\label{eom3}
M(0) \frac{d^2 \Delta q(t) }{dt^2} + \gamma(0) \frac{d \Delta q(t)}{dt} + C(0) \Delta q(t) = 0 \ ,
\end{equation}
where the mass and friction coefficients are defined as
\begin{equation}\label{emh}
M(0)=\frac{1}{2}  \left. \frac{d^2 {\chi}(\omega)}{d \omega^2} \right|_{\omega=0}, \quad \gamma(0)= -i \left. \frac{d {\chi}}{d \omega} \right|_{\omega=0}.
\end{equation}
By derivation of (\ref{eom3}) we used the relation (\ref{couple2}) between stiffness and coupling constant, and we kept in mind that, within the harmonic approximation, the  $\Delta q(t)$ is proportional to $ \exp{(-i\omega t)}$, so that $d \Delta q(t)/dt\propto -i\omega \Delta q(t) $.

In case that the mean field Hamiltonian depends on few collective coordinates, $q=\{q_{\mu}\}$, the expansion (\ref{expand}) should be generalized to
\begin{eqnarray}\label{expand2}
\hat H_{mf}(q)=\hat H_{mf}(q_0)+\sum_{\mu}(q_{\mu}-q_{\mu}^{(0)})\frac{\partial\hat H_{mf}}{\partial q_{\mu}^{(0)}}\nonumber\\
+\frac{1}{2}\sum_{\mu\nu}(q_{\mu}-q_{\mu}^{(0)})(q_{\nu}-q_{\nu}^{(0)})\left<\frac{\partial^2\hat H_{mf}}{\partial q_{\mu}^{(0)}\partial q_{\nu}^{(0)}}\right>.
\end{eqnarray}
Correspondingly, the response functions are modified to
\begin{eqnarray}\label{chitmunu}
\widetilde\chi_{\mu\nu}(t-s)=\Theta(t-s)\frac{i}{\hbar}\langle [\hat F^I_{\mu}(t),\hat F^I_{\nu}(s)] \rangle_{q_0}\ ,\nonumber\\
\chi_{\mu\nu}(\omega) = \sum_{jk}  \frac{n_k - n_j}{\hbar \omega - \left(\varepsilon_k - \varepsilon_j \right) + i\epsilon}F_{\mu}^{jk} F_{\nu}^{kj}\ ,
\end{eqnarray}
and the mass and friction coefficients turn into mass and friction tensors
\begin{equation}\label{zerolim}
M_{\mu\nu}(0)=\frac{1}{2}  \left. \frac{d^2 {\chi_{\mu\nu}}(\omega)}{d \omega^2} \right|_{\omega=0}, \, \gamma_{\mu\nu}(0)= -i \left. \frac{d {\chi_{\mu\nu}}}{d \omega} \right|_{\omega=0}\,.
\end{equation}
In the case of absence of pairing effects, the tensors of friction and mass can be calculated directly by differentiating $\chi_{\mu\nu}(\omega)$ (\ref{zerolim}) with respect to $\omega$.
In the presence of pairing effects it was suggested in \cite{ivahof} to use as $\hat H_{mf}$ the independent quasi particles Hamiltonian
\begin{eqnarray}\label{hbcs}
\hat H_{BCS}&=&\sum_k2v_k^2 (\varepsilon_k-\lambda)-2\Delta\sum_ku_kv_k+\frac{\Delta^2}{G}\nonumber\\
&+&\sum_k E_k(\alpha_k^{\dagger}\alpha_k+\alpha_{\bar k}^{\dagger}\alpha_{\bar k}),
\end{eqnarray}
where $\lambda$ is the chemical potential, $G$ and $\Delta$ are the pairing strength constant and the pairing gap respectively, $u_k$ and $v_k$ are the coefficients of the Bogoliubov-Valatin transformation and $E_k$ are the quasiparticle energies,
\begin{equation}\label{ekbcs}
E_k=\sqrt{(\varepsilon_k-\lambda)^2+\Delta^2},
\end{equation}
and $\alpha_{\bar k}^{\dagger}$ and $\alpha_{\bar k}$ are the creation and annihilation operators for the quasi-particles. The operator $\hat F$ in
quasi-particle representation has the form
\begin{eqnarray}\label{hatf}
\hat{F}&=&\sum_k F_{kk} 2 \upsilon_k^2 + \sum_{jk} F_{kj} \xi_{kj} \left( \alpha_k^{\dag} \alpha_j + \alpha_{\bar{k}}^{\dag} \alpha_{\bar{j}} \right) \nonumber\\
&+& \sum_{kj} F_{kj} \eta_{kj} \left( \alpha_k^{\dag} \alpha_{\bar{j}}^{\dag} + \alpha_{\bar{j}} \alpha_k \right).
\end{eqnarray}

Inserting (\ref{hatf}) into (\ref{chitmunu}) after somewhat lengthy but
straightforward calculation one gets
\begin{eqnarray}\label{chitBCS}
&&\widetilde\chi_{\mu\nu}(t)=
\frac{-2\Theta(t)}{\hbar}\sum_{kj}^{\prime}(n_k^T-n_j^T)\xi_{kj}^2 F_{\mu}^{jk} F_{\nu}^{kj}
\sin(E_{kj}^-t/\hbar)\nonumber\\
&&-\frac{2\Theta(t)}{\hbar}\sum_{kj} (n_k^T+n_j^T-1)\eta_{kj}^2 F_{\mu}^{jk} F_{\nu}^{kj}
\sin(E_{kj}^+t/\hbar),
\end{eqnarray}
where
$E_{kj}^-\equiv E_k-E_j\,, E_{kj}^+\equiv E_k+E_j\,,
\eta_{kj} \equiv u_k \upsilon_j + \upsilon_k u_j \,, \xi_{kj} \equiv u_k u_j - \upsilon_k \upsilon_j\,,$
and the temperature dependent occupation numbers are defined as
\begin{equation}
n_k^T = 1 / (1+e^{{E_k}/{T}}).
\end{equation}
The diagonal components of $\xi$ term of operator $\hat F$ commute
with Hamiltonian (\ref{hbcs}) and thus do not contribute to response
function (\ref{chitBCS}). That is why the first sum in (\ref{chitBCS}) is marked
by a prime. In the second sum of (\ref{chitBCS}) both diagonal and non-diagonal
components contribute.
The Fourier transform of (\ref{chitBCS}) leads to
\begin{eqnarray}\label{chiomlong}
\chi_{\mu\nu}(\omega)=\sum_{jk}^{\prime}
\frac{(n_k^T-n_j^T)\xi_{kj}^2}{\hbar\omega-E_{kj}^-+i\epsilon}F_{\mu}^{jk} F_{\nu}^{kj} \nonumber\\
+\sum_{jk}
\frac{(n_k^T+n_j^T-1)\eta_{kj}^2}{\hbar\omega -E_{kj}^++i\epsilon}F_{\mu}^{jk} F_{\nu}^{kj} \,,
\end{eqnarray}
and the tensors of friction and mass (\ref{zerolim}) turn into
\begin{align}\label{micro_fric}
\gamma_{\mu \nu} (0) &= 2 \hbar \sum_{jk}^{\prime} (n_k^T - n_j^T ) \xi_{kj}^2 \frac{ E_{kj}^-  \Gamma_{kj}}{\left[ ( E_{kj}^-)^2 + \Gamma_{kj}^2 \right]^2 } F^{kj}_\mu F^{jk}_{\nu} \nonumber \\
+& 2 \hbar \sum_{jk}  \left(n_k^T + n_j^T -1 \right) \eta_{kj}^2 \frac{ E_{kj}^+  \Gamma_{kj} }{[ ( E_{kj}^+ )^2 + \Gamma_{kj}^2 ]^2} F^{kj}_\mu F^{jk}_{\nu} \,,
\end{align}
\begin{align}
\label{micro_mass}
M_{\mu\nu}(0)=\hbar^2\sum_{jk}^{\prime}(n_k^T - n_j^T)\xi_{kj}^2\frac{E_{kj}^{-2} [E_{kj}^--3\Gamma_{kj}]}{[( E_{kj}^-)^2+\Gamma_{kj}^2]^3} F^{kj}_\mu F^{jk}_{\nu}  \nonumber \\
+\hbar^2 \sum_{jk}  \left(n_k^T + n_j^T -1 \right) \eta_{kj}^2 \frac{E_{kj}^{+2}[E_{kj}^+-3 \Gamma_{kj}]}{[( E_{kj}^+)^2+\Gamma_{kj}^2]^3} F^{kj}_\mu F^{jk}_{\nu}  \ .
\end{align}
In the above expressions for friction and mass tensors the infinitely small quantity $\epsilon$ was replaced by the average of the collisional widths $\Gamma_k$ and $\Gamma_j$ of $k$ and $j$ states, $\Gamma_{kj}=(\Gamma_k+\Gamma_j)/2$. In this way one can take into account the effect of residual interaction $\hat V^{(2)}_{res}$, which is absent in the mean-field Hamiltonian.
The calculation of $\Gamma_k$ is discussed in detail in \cite{ivahof}.

For the description of the fission process we will solve the Langevin equation for the time evolution of parameters which define the shape of the nuclear surface. The shape of the nucleus in the present calculations is parametrized by the shape parametrization of the two-center shell model with three deformation parameters. So, we will need multidimensional tensors of friction and mass.
For this purpose we will use expressions (\ref{micro_fric}) and (\ref{micro_mass}) obtained within the linear response approach.
These expressions were derived within the quantum approach with the shell and pairing effects taking into account. In what follows we will call expressions (\ref{micro_fric}) and (\ref{micro_mass}) {\it microscopic transport coefficients}.  

The comparison of the microscopic and macroscopic transport coefficients calculated within the two-center shell model shape parametrization can be found in \cite{jnst2016,usang2016}.
 At large temperatures both microscopic friction and mass coefficients look similar to macroscopic friction and mass coefficients. At small temperatures the microscopic and macroscopic transport coefficients deviate from each other very much. The microscopic mass tensor  decreases with increasing temperature $T$, while the friction tensor increases as $T$ increases, and the macroscopic mass and inertia tensors are temperature independent. Thus, the results of dynamical calculations at low excitation energies with the microscopic and macroscopic transport coefficients can deviate from each other, since at the saddle the temperature can be quite small.



\section{The calculated results}
\label{results}
We start the integration of the Langevin equations \eqref{langevin} from the initial point $q_0=\{z_0/R_0=0.8,\,\delta=0.2,\,\alpha=0.0\}$. This point corresponds to the second minimum on the potential energy surface. The evolution of $q(t)$ over time will generate a trajectory across the potential surface. On each step of integration the neck radius is checked. If the trajectory reaches a scission point, $r_{neck}=0$, we consider such a trajectory to be a fission event. Each fission event provides  information on the time it took for scission, the value of collective coordinates and velocities, the prescission kinetic energies and the intrinsic excitation energy (temperature) at scission. Due to the presence of the random force, each trajectory gives somewhat different results. The integration of Eq. \eqref{langevin} is repeated typically up to 500 000 times in order to get results that are stable with respect to the number of trajectories.

\subsection{Mass distribution of fission fragments}
\label{yields}
In this section, we compare the calculated mass distributions with experimental information.  The upper limit of the excitation energy was
set to 20 MeV.  We are aware of the fact that there are contributions from second and third chance fission above several MeV.  The effect of multi chance fission may be estimated by using the Hauser-Feshbach model calculation with generically available codes such as TALYS \cite{talys} or EMPIRE \cite{empire}.  Still, we compare here the
calculated results for only first chance fission with experimental data, since it is not our purpose to fit the data with our model; rather, we wish to understand the reaction mechanisms in terms of the fluctuation dissipation dynamics.
\begin{figure}[b]
\includegraphics[width=0.4\textwidth]{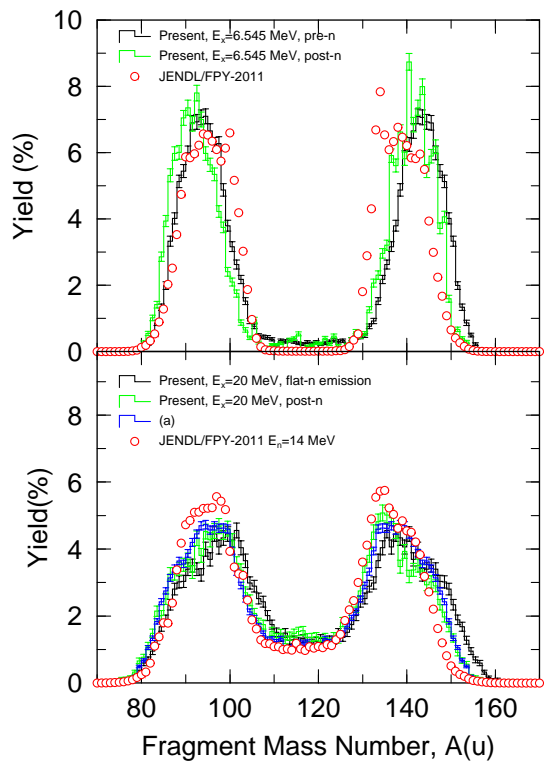}
\caption{Fission fragment yield of $^{236}$U. Top: at $E_x$=6.545 MeV, bottom : at $E_x$=20 MeV. The black histogram shows our calculation with pre-neutron (pre-n) emission for $E_x=6.545$ MeV and flat-neutron (flat-n) emission for $E_x=20$ MeV.
The green histogram shows our calculation with neutron emission calculated from \cite{muller1984} (top) and neutron emission calculated using GEF \cite{gef} (bottom).
Our calculations (denoted by red circles) are compared with evaluated post-neutron (post-n) distributions stored in JENDL.
We also compare them with (a) our previous data \cite{usang2016} for single-chance, flat-neutron (flat-n) emission calculations using microscopic transport coefficients.}\label{fig2}
\end{figure}

We obtain the mass distributions of fission fragments  from the number of trajectories with given $\alpha$ that managed to reach the scission configuration. For positive value of $\alpha$, the light fragment mass number is expressed as $A_L=(A/2)(1-\alpha)$, and for the heavy fragment it is $A_H=(A/2)(1+\alpha)$. For negative value of $\alpha$, the converse is true.


The application of effective temperature and microscopic transport coefficients allowed us to obtain using the 3D Langevin calculation quite reasonable mass distribution height and width for the $^{236}$U compound nucleus; see Fig. \ref{fig2}. The excitation energy $E_x=6.545$ MeV corresponds to fission fragment yield from thermal incident neutron data. Hence, comparison is made with the evaluated thermal incident neutron fission product yield, such as that from JENDL 4.0 \cite{JENDL}. The calculated fission fragment light and heavy mass averages are $\left<A_L\right>=93.92$ and $\left<A_H\right>=141.83$, while the JENDL fission product mass averages are 94.75 and 138.68. We expect our fission fragment yield to deviate slightly from the fission product yield from evaluated data because prompt neutron emissions are not included. Deviations of our light mass averages from JENDL are quite reasonable, but deviations from JENDL heavy mass averages are quite large. However if we look through the perspective of Flynn's mass average systematics \cite{flynn1972}, the deviation of our results from Flynn's $\left<A_H\right>_{\mathrm{Flynn}}=139\pm1$ is still acceptable.

At $E_x=20$ MeV the mass averages in our microscopic calculations with effective temperature also seems to be shifted slightly  towards heavier mass in comparison to our previous calculations \cite{usang2016} that did not use the effective temperature treatment. Both calculations are for the first chance fission calculation and a flat neutron emission are assumed for fair comparison with previous data. The mass averages of the fragments are $\left<A_L\right>=97.01$ and $\left<A_H\right>=136.94$, respectively; a little off from evaluated fission fragment mass averages for a 14 MeV neutron incident on $^{235}$U, but still close to the expected fission fragment mass average systematics.
\begin{figure}[t]
\includegraphics[width=0.4\textwidth]{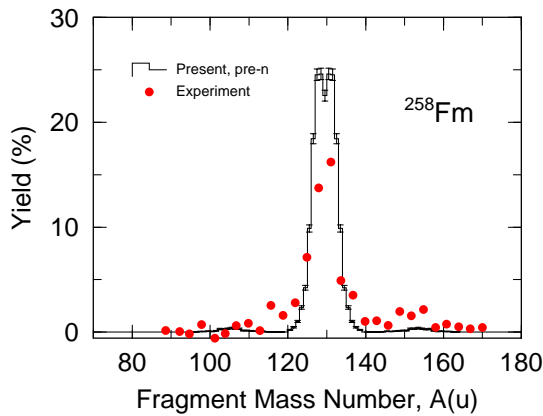}
\caption{Fission fragment yield for $^{258}$Fm at $E_x=3.34$ MeV in comparison with experiment \cite{hoffman1980}.}\label{fig3}
\end{figure}

Using the same methodology, we made an attempt to describe the fission of $^{258}$Fm at an excitation energy of 3.34 MeV. For the fragment mass distribution we got a single symmetric peak with mass average around $A_F=$129. Our calculations are very close to the experimental results \cite{hoffman1980} (the fission yield peak is at $A_F=$130) shown in Fig. \ref{fig3}.

\begin{figure}[t]
\includegraphics[width=0.4\textwidth]{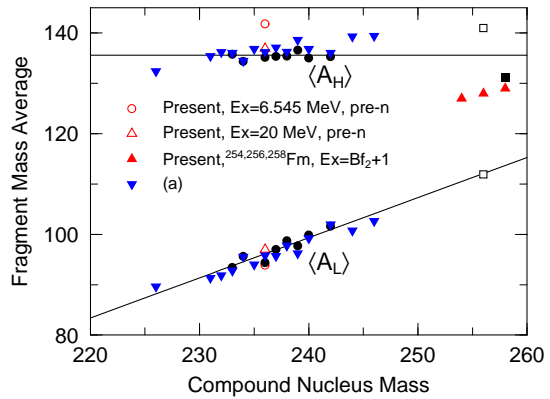}
\caption{\label{systematic1} 
Fission fragments mass systematics (solid line) obtained by fitting the average mass for light and heavy fission fragments (filled circles). These mass averages are calculated from the fission fragment mass yield of JENDL 14 MeV incident neutron data. We also plot $^{256,258}$Fm experimental results from \cite{flynn1972} (open squares) and \cite{hoffman1980} (filled squares). We compare them with  (a) our previous data \cite{nd2016,ines5} (blue inverted triangles) for compound nucleus at $E_x=20$ MeV.
}
\label{fig4}
\end{figure}

In Fig. \ref{fig4} we collected the results of present calculations for $^{236}$U at $E_x$=6.545 and 20 MeV, and for $^{254,256,258}$Fm at 1 MeV above the second fission barrier $B_{f2}$, together with the data from our previous calculations \cite{nd2016,ines5} and experimental mass averages.  The barriers $B_{f2}$ are obtained from generic calculations using the GEF code \cite{gef}. Thus, the $^{254,256,258}$Fm isotopes are calculated at $E_x$ of 4.07, 3.66 and 3.34 MeV respectively.
GEF is used in the estimation of $B_{f2}$ as it is computationally fast and the barrier heights are usually adjusted to the experimental values.

At present, however, we are still unable to reproduce the transition from the double peak in $^{256}$Fm to the single peak in $^{258}$Fm by 3D Langevin calculations. All Fm isotopes in the current calculations have a single peak mass yield. We see also some traces of events with standard fission modes.
\subsection{The total kinetic energy}
\label{TKE}
We calculate TKE as a sum of prescission kinetic energy in the fission direction, KE$_{pre}$, and the Coulomb repulsion energy KE$_{Coul}$.
The Coulomb repulsion energy is calculated using the point charge approximation for the sake of computational speed and simplicity;
\begin{equation}
\mathrm{KE}_{\mathrm{Coul}}=e^2\frac{Z_1 Z_2}{R_{12}}\,\, \quad \text {with}\quad e^2=1.44\, \text {MeV} \text{fm},
\end{equation}
where $R_{12}$ is the distance between centers of mass of left and right parts of the nucleus at the scission point.

In the phenomenological treatment of TKE profiles \cite{brosa1990} the contributions of standard, super-long, and super-short fission modes are mentioned and related to the prescission shape of the nucleus.
As the name of these fission modes indicate, the super-short, standard, and super-long fission modes correspond to a short, medium and long elongations, respectively.
It was suggested that these fission modes came from different fission channels along different paths on the potential energy surface. 

\begin{figure}[t]
\includegraphics[width=0.4\textwidth]{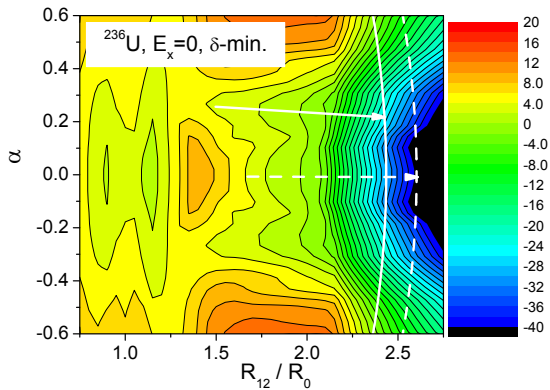}
\caption{The minimized-in-$\delta$ potential energy surface of $^{236}$U at $T=0$ calculated within TCSM. The white line shows the position of zero neck radius for $\delta$=0 (solid) and $\delta$=0.1 (dash).}
\label{fig5}
\end{figure}

This suggestion was based on the results of fitting the mass distributions by a few Gaussians. The existence of super-short, standard, and super-long fission valleys was not confirmed by the calculations of potential energy surfaces.
In the cases of uranium, thorium, and californium most of the fission events came from the standard fission mode. On the potential energy surface of $^{236}$U (see Fig.~\ref{fig5}), one can clearly see the fission valley of the standard mode.
\begin{figure}[b]
\includegraphics[width=0.4\textwidth]{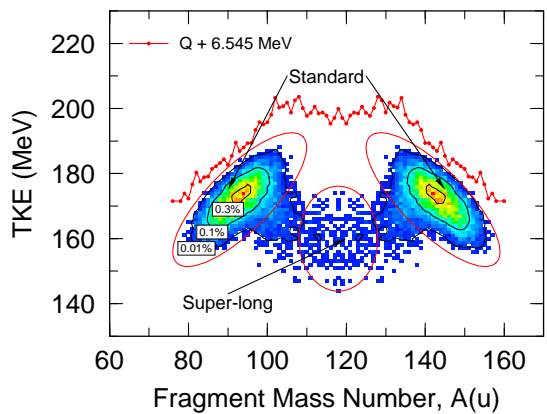}
\caption{The calculated distribution of fission events of $^{236}$U in kinetic energy and fragment mass  at $E_x=6.545$ MeV. The red curve denotes the
kinematically allowed maximal value of TKE, namely, $Q+E_x$.}
\label{fig6}
\end{figure}
From the numerical results it follows that the main contribution to the standard mode comes from shapes with $\delta\approx 0$. The scission line for $\delta=0$ is shown in Fig.~\ref{fig5} by a white solid line.
There is also some hint of another valley at mass symmetric deformations, $\alpha\approx 0$. This valley is caused mainly by somewhat longer shapes, $\delta\approx 0.1$. In this sense the second valley can be referred to as super-long.

The contributions of both standard and super-long, modes are clearly seen in the mass-energy distribution of fission fragments; see Fig.~\ref{fig6}.
As one could expect, the calculations of TKE  show a strong standard fission mode. Around symmetric splitting there is also a small contribution from the super-long fission mode.

\begin{figure}[t]
\includegraphics[width=0.4\textwidth]{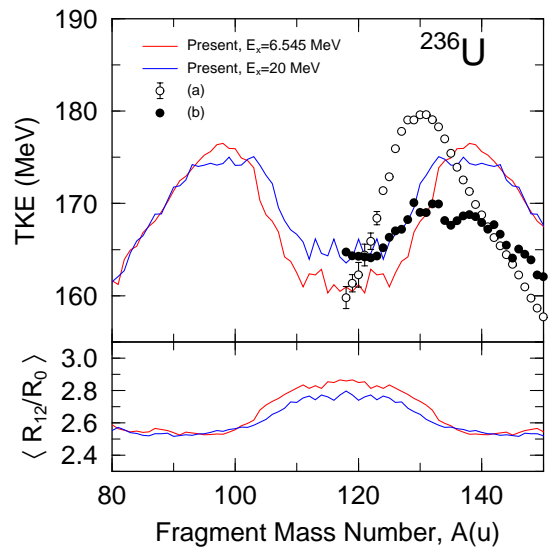}
\caption{The average value of TKE (top) and average elongation $<R_{12}>$ of fissioning nucleus at the scission point for $E_x=6.545$ MeV (red) and $E_x=20$ MeV (blue). The experimental data are taken from (a) \protect\cite{zeynalov} ($E_n\approx 0$) and (b) \protect\cite{exdepend_u235} ($E_n=$ 15.5 MeV), correspondingly.}
\label{fig7}
\end{figure}

The comparison of calculated TKE distributions for $^{236}U$ at $E_x=$ 6.545 MeV and $E_x=$ 20 MeV with experimental data is shown in Fig.~\ref{fig7}.
In the bottom part of Fig.~\ref{fig7} we show the mean elongation of the nucleus at the scission point at fixed mass asymmetry: the average value of the distance between centers of mass of left and right parts of the nucleus,
\begin{equation}\label{R12avr}
<R_{12}(A)> =\sum_{i=1}^{N_A} R_{12}(A_i,r_{neck}=0)\,\bigg /N_A\,.
\end{equation}
The summation in (\ref{R12avr}) is carried out over the trajectories $i$ with the fragment mass $A_i$ that fulfills the condition $A-1/2 \leq A_i \leq A+1/2$ and the sum of these trajectories is $N_A$.
From the bottom part of Fig. \ref{fig7} one can see that for larger excitation energy the scission shapes around symmetric splitting (super-long mode) become somewhat shorter. Consequently, the Coulomb repulsion energy and fission fragment kinetic energy become larger. 

The decrease of TKE at symmetric splitting is in accord with the experimental results, but the total agreement of calculated and measured distributions of TKE is not so good. The reason could be the restricted (three-dimensional) shape parametrization. We tried to do the calculations with $\delta_1\neq\delta_2$. The preliminary results \cite{submitted} show that 4D Langevin calculations reproduce experimental TKE distributions much more accurately. Unfortunately, 4D Langevin calculations are much more time consuming.

\begin{figure}[t]
\includegraphics[width=0.4\textwidth]{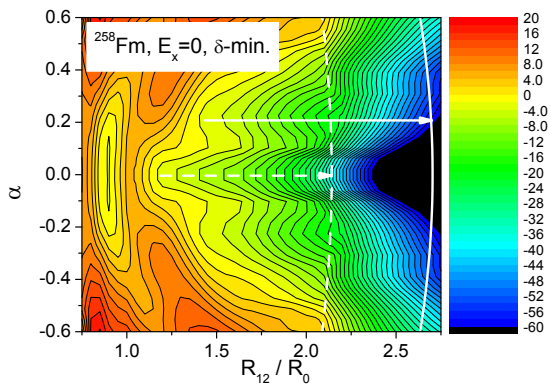}
\caption{The minimized-in-$\delta$ potential energy surface of $^{258}$Fm at $T=0$ calculated within TCSM. The white line shows the position of zero neck radius for $\delta$=-0.2 (dash) and $\delta$=0.15 (solid).}
\label{fig8}
\end{figure}

In the case of $^{258}$Fm, we could infer from Fig.~\ref{fig8} that the standard fission mode also exists but it is not a dominant fission channel. The $^{258}$Fm fission is predominantly mass symmetric. The fission fragments in this case are close to the double magic $^{132}$Sn. Due to the very strong shell structure in spherical $^{132}$Sn, the configuration just before fission consists of two almost spherical fragments. Within the TCSM shape parametrization, we can get almost spherical fragments with {\it negative} $\delta$. Such a configuration is very short, see Fig.~\ref{fig8}. So, the corresponding fission valley can be referred to as super-short. The main contribution to the TKE in the case of $^{258}$Fm comes from the super-short mode with large kinetic energy.

\begin{figure}[b]
\includegraphics[width=0.4\textwidth]{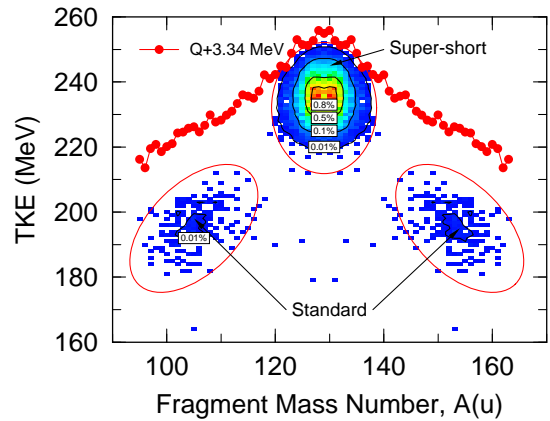}
\caption{The calculated mass-energy distribution of fission events of $^{258}$Fm at $E_x=3.34$ MeV.  The red curve denotes the kinematically allowed maximum values of TKE, namely, $Q+E_x$.}
\label{fig9}
\end{figure}

Another (smaller) contribution seen in Figs.~\ref{fig8} and \ref{fig9} can be referred as the standard mode. It corresponds to more elongated mass asymmetric scission shapes with lower kinetic energy.

\begin{figure}[b]
\includegraphics[width=0.49\textwidth]{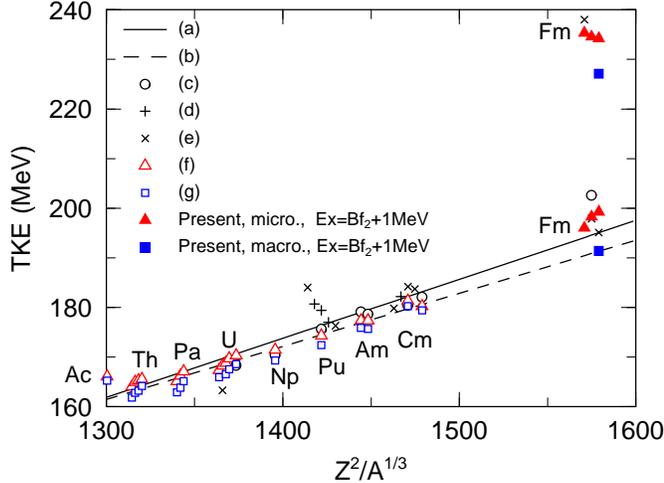}
\caption{Systematical trends of average TKE of fission fragments as a function of $Z^2/A^{1/3}$ of the fissioning system. We compare our present results with (a) \cite{viola1985} and (b) \cite{viola1966} linear least square of the TKE systematics as a function of the fissioning system. Included are (c) evaluated data from JENDL \cite{JENDL} and experimental spontaneous TKE from (d) \cite{hoffman1995} and (e) \cite{hoffman1996}. We compare also with several other nuclides using our previous results at $E_x=20$ MeV \cite{nd2016,ines5} for (f) microscopic and (g) macroscopic transport coefficients.}
\label{fig10}
\end{figure}

We are positive in our assessments that these traces  are the standard fission mode because it has a $\left<\mathrm{TKE}\right>=198.29$ MeV, which is close to $\left<\mathrm{TKE}\right>$ values from $^{256}$Fm that are dominated by standard fission modes as we can see from the prominent two peaks in its fission yield. The dominant fission modes in $^{258}$Fm, however, are obviously due to the super-short fission modes that have $\left<\mathrm{TKE}\right>=234.58$ MeV. This helps us show that Viola's systematics \cite{viola1985} only tracks the TKE due to standard fission modes as is clear from Fig.~\ref{fig10}.


The peculiar emergence of the standard modes, however, could be seen in the TKE profiles only if the effective temperature formulation is in use. We know from experimental evidence \cite{hulet1989} that for $^{258}$Fm the two fission modes coexist.
We start calculations from the mass symmetric minimum in the potential energy surface, and almost all trajectories immediately fall down into a very deep super-short fission valley. Only if the fluctuations are very large can some trajectories jump into the standard fission valley, which lay much higher in energy.

This crossover to another valley is possible only because the effective temperature allowed the trajectories to experience stronger random fluctuations. We can see this from Eq. \eqref{Einstein relation}. For example, given $T=0.5$ MeV from \eqref{tstar} we will find for $T^*$ the saturated value of 1 MeV. This means that $g_{ij}$ is $\sqrt{2}$ times larger in magnitude, giving it the necessary impetus to cross to the nearby valley.

Figure \ref{fig10} exhibits a comparison of the average TKE values calculated by our 3D Langevin model with experimental data, Viola's systematics \cite{viola1985}, and values given in the evaluated library JENDL 4.0.  From Ac to Cm, both the experimental and calculated results are in accord with the monotonically increasing trends given by Viola's systematics.  However, there are two groups in Fm isotopes: a lower TKE group which agrees well with Viola's systematics, and a higher energy one which corresponds to the super-short fission mode.  Our calculation with microscopic transport coefficients (shown by triangles) agrees better than those of macroscopic transport coefficients to both the Viola's trends for lower TKE group and the abnormally high TKE group for Fm region.

\subsubsection{Average TKE dependence on excitation energy}

\begin{figure}[t]
\includegraphics[width=0.4\textwidth]{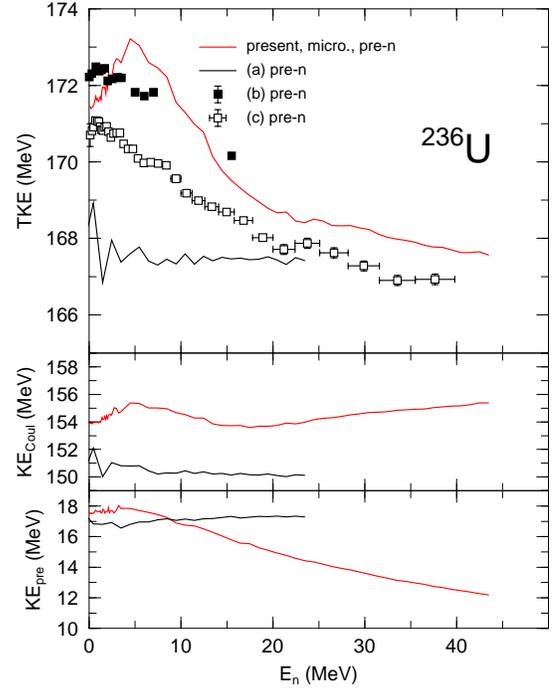}
\caption{The TKE as a function of neutron energy $E_n$ calculated for compound nucleus $^{236}$U. The red lines are the present results for pre-neutron (pre-n) emission, single chance fission TKE, KE$_{pre}$, and KE$_{Coul}$. We compare $^{236}$U with; (a) our previous data \cite{nd2016} calculations with macroscopic transport coefficients without using the effective temperature description in Eq. \eqref{Einstein relation}, (b) experimental data from \cite{exdepend_u235}, and (c) experimental data from \cite{duke2015phd}.}\label{fig11}
\end{figure}

\begin{figure}[t]
\includegraphics[width=0.4\textwidth]{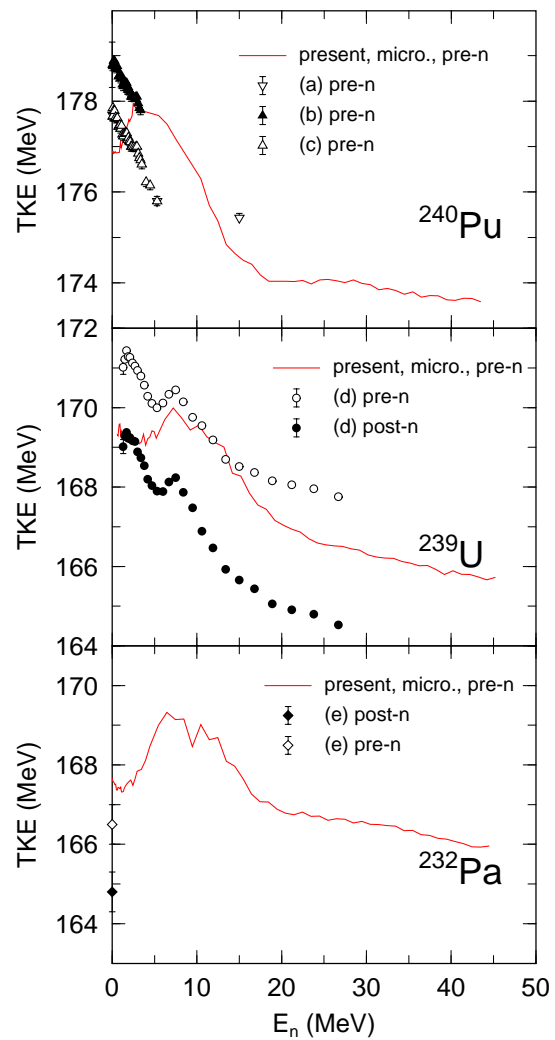}
\caption{The TKE as a function of neutron energy $E_n$ calculated for three compound nuclei; $^{239}$U, $^{240}$Pu, and $^{232}$Pa on the top, middle, and bottom panels respectively. The red lines are the present results for pre-neutron emission (pre-n), single chance fission for the corresponding compound nucleus TKE. 
Our calculated TKE's for $^{240}$Pu are compared with experimental data from  (a) \cite{surin1972}, (b) \cite{vorobeva1974}, and (c) \cite{Akimov1971}. 
We compare the $^{239}$U fission TKE from post-neutron (post-n) and pre-neutron emission with (d) experimental data from \cite{duke2016}. 
$^{232}$Pa TKE's are compared with experimental data for thermal neutrons in (e) \cite{asghar1978}. }\label{fig11-2}
\end{figure}

In the present work we have also calculated and compared with experimental data the dependence of average TKE values on the kinetic energy $E_n$ of incident neutrons in Fig. \ref{fig11} and \ref{fig11-2}. For this we related $E_n$ to the excitation energy $E_x$ by the formula $E_n=E_x-S_n$. The neutron separation energies were calculated using
$S_n=M_{A-1}+M_n-M_A$, where $M_{A-1}$  is the target mass excess, $M_{A}$ is the compound mass excess, and $M_n$ is the mass excess of a neutron.  The values of mass excess was obtained from Reference Input Parameter Library (RIPL-3) that gave either experimental or recommended mass data given by \cite{audi-mass}. We only fall back to the theoretical mass data in RIPL-3 in the very rare cases when no experimental or recommended values are available.

The separation energy values we obtained are 6.545, 4.806, 6.534, and 5.549 MeV for $^{236}$U, $^{239}$U, $^{240}$Pu, and $^{232}$Pa respectively. We assumed that, in the case of thermal neutrons, the excitation energy is approximately equal to the neutron separation energy. The compound nuclei produced in the reaction of neutrons with fissile nuclei such as $^{236}$U, $^{240}$Pu, and $^{232}$Pa have a fission barrier lower than the neutron separation energy, so the nucleus can easily undergo fission and we could  obtain sufficient statistics from incident neutrons at thermal energy.
In the case of the compound nucleus $^{239}$U, the neutron separation energy is lower than the fission barrier. 
The compound nucleus $^{239}$U could be produced by bombarding the fertile $^{238}$U with neutrons.
Fission could only occur when the excitation energy used in our calculation was higher than the fission barrier.
Since sufficient statistics are obtained at $E_n=500$ keV, it means that the fission barrier of $^{239}$U (approximately equal to 5.30 MeV) was overcome.

As depicted in Fig. \ref{fig11}, in the case of $^{236}$U the average TKE calculated by Langevin procedure with microscopic transport coefficients increase from 171 to 173 MeV for neutron energies $0\leq E_n \leq 5$ MeV, and then decays to 167 MeV.
Experimentally observed average TKE values \cite{duke2015phd,kuzminov} increase slightly for $0 \leq E_n < 2$ MeV and afterwards decrease at higher $E_n$. Such an increasing trend of TKE at low $E_n$ values is reproduced, although not perfectly, by the present Langevin calculation with microscopic transport coefficients. 
In contrast, results from our Langevin calculation with macroscopic transport coefficients \cite{nd2016} are approximately constant, giving an average TKE of 167.5 MeV irrespective of $E_n$. 
We can see that in comparison to results with macroscopic transport coefficients, those with the microscopic transport coefficients are much closer to the trend of experimental pre-neutron data and behave similarly with higher $E_n$.  Therefore it is important to employ the microscopic transport coefficients.

As one can see in the middle panel of Fig.~\ref{fig11}, the Coulomb repulsion energy practically does not depend on $E_n$.
This means that increase of the super-long mode, which gives lower Coulomb repulsion energy, is not the origin of the decrease of the TKE as excitation energy increases.
The decrease of the TKE, is, thus, brought by the decrease of the prescission kinetic energy KE$_{pre}$.
This effect can be easily understood. 
In the microscopic approach the friction force is larger for higher excitation energies. Consequently, the motion in the fission direction gets slower and the  prescission kinetic energy turns out to be smaller. We have checked that 3D Langevin calculations with {\it macroscopic} transport coefficients do not show any dependence of TKE on the excitation energy $E_x$.

We also plotted the average TKE for $^{239}$U, $^{240}$Pu and $^{232}$Pa as a function $E_n$ in Fig. \ref{fig11-2}.
In the case of $^{239}$U, the average TKE increase from 169 MeV at $E_n=0.5$ MeV to 170 MeV at $E_n \approx 8$ MeV and then decays to 166 MeV at $E_n\approx 45$ MeV.  The calculated TKE values behave similarly to the pre-neutron and post-neutron experimental data \cite{duke2016} and have average TKE values between pre-neutron and post-neutron experimental values.

Experimental TKE values for $^{240}$Pu at $E_n$ up to around 5 MeV \cite{surin1972,vorobeva1974,Akimov1971} do not increase as in $^{236,239}$U but linearly decrease with $E_n$. This is contrary to the pattern we see from our calculation but the absolute values of the calculated TKE are close to experimental values. Experimental values for $E_n > 5$ MeV \cite{surin1972} seems to saturate at 175 MeV. The calculated TKE also decays at $E_n$ higher than 5 MeV and then saturates at 174 MeV.

There are insufficient experimental data to study change of average TKE with $E_n$ for $^{232}$Pa. The calculated average TKE increases from 167 to 169 MeV and then decrease to 166 MeV. The only experimental data are pre-neutron average TKE values from thermal neutrons \cite{asghar1978}.
In all the cases, the TKE decreases as $E_n$ (therefore excitation energy) increases.  Our calculations follow this general trend.

\begin{figure}[t]
\centering
\includegraphics[width=0.45\textwidth]{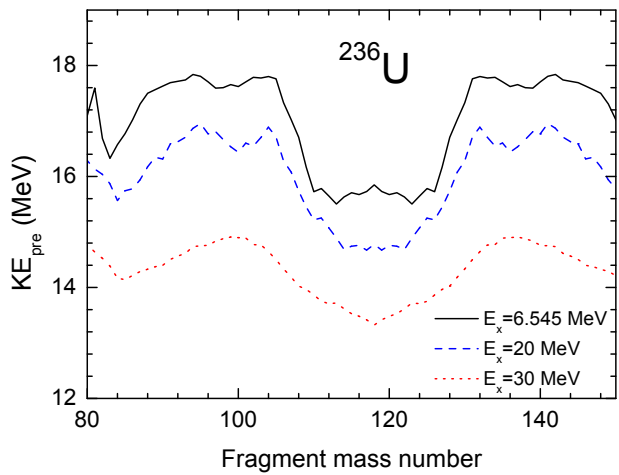}
\caption{The dependence of average KE$_{pre}$ on the fragment mass. KE$_{pre}$ was calculated for  $E_x=6.545$ MeV (black), $E_x=20$ MeV (blue), and $E_x=30$ MeV (red).   }
\label{fig12}
\end{figure}

In Fig. \ref{fig12},  we plotted the distributions of prescission kinetic energies KE$_{pre}$ for $^{236}$U as functions of fragment mass at excitation energies of 6.545, 20, and 30 MeV. As we can see, the average KE$_{pre}$ for the so-called super-long mode, which stays around the region of symmetric mass division, does not decrease much as excitation energy increases.  
Instead, the average KE$_{pre}$ for the standard modes gradually decreases when excitation energy increases.  
This is the reason for the decrease of total average TKE as excitation energy increases.
This happens because, as excitation energy (therefore temperature) increases, the microscopic friction tensors generally increase, which results in less momentum gain during the descent from the saddle to scission configurations.
This again shows the importance of using the microscopic transport coefficients which are dependent on the temperature of the system instead of using the traditional macroscopic ones. This tendency is, of course, somewhat diluted if we consider the effects of multi-chance fission.

\section{Summary}
\label{summa}

We have calculated the mass distribution and the total kinetic energy of fission fragments for a series of actinides and Fm isotopes at various excitation energies within the three-dimensional Langevin approach with microscopic transport coefficients. For the diffusion tensor we used the modified Einstein relation with an effective temperature that accounts for the quantum features of the fluctuation-dissipation theorem.  

The systematic trends of TKE as a function of both
$Z^2/A^{1/3}$ of the fissioning system and excitation energies are well reproduced by the present calculations.  The sudden appearance of the super-short mode in the Fm region is also well reproduced.  It was found that the decrease of the average TKE with growing excitation energy is due to the decrease of the prescission kinetic energy, not by the Coulomb repulsion energy. The decrease of prescission kinetic energy has a clear reason: the microscopic friction tensor gets larger for larger excitation energy. Consequently, the collective motion become slower and the kinetic energy smaller. This demonstrates the importance of the dynamical description of the fission process and the use of the microscopic transport coefficients.

Without introduction of effective temperature, it is difficult for us to probe the fission reaction of $^{236}$U, $^{239}$U, $^{240}$Pu, and $^{232}$Pa at low excitation energy. As a bonus, we found that it is a necessary ingredient to see the emergence of standard fission modes in $^{258}$Fm.
We realize that using a more accurate prescription for the effective temperature could give  better results but in this case it might be necessary to adjust it for different nuclides.

In the present calculations, the mass yields are not perfect but they still give reasonable values.
For a better description one should introduce a more flexible (4D) shape parametrization and account for the contributions from the multi-chance fission.
After all, it is difficult to reproduce the evaluated fission product yield without reproducing a good prompt neutron emission multiplicity. 
An attempt to describe the prompt neutron emission from its charge polarizations as was undertaken in \cite{ishizuka-ines5}.

\begin{acknowledgments}
The present study includes the results of "\textit{Comprehensive study of delayed-neutron yields for accurate evaluation of kinetics of high-burn up reactors}" entrusted to Tokyo Institute of Technology by the Ministry of Education, Culture, Sports, Science and Technology of Japan (MEXT).
We would also like to acknowledge the IAEA CRP on beta-delayed neutrons (F41030).
M.D.U would like to thank Hitachi-GE for the scholarship received during the period of this study.
\end{acknowledgments}

\bibliography{stdref.bib}
\end{document}